\documentclass{article}
\usepackage{arxiv}
\usepackage{amssymb}
\usepackage{amsthm}
\usepackage[english]{babel}
\usepackage{hyperref}
\usepackage[utf8]{inputenc}
\usepackage{tikz}
\usetikzlibrary{arrows, calc, shapes, backgrounds,decorations.pathmorphing,patterns}
\pgfdeclaredecoration{penciline}{initial}{
    \state{initial}[width=+\pgfdecoratedinputsegmentremainingdistance,
    auto corner on length=1mm,]{
        \pgfpathcurveto%
        {
            \pgfqpoint{\pgfdecoratedinputsegmentremainingdistance}
                      {\pgfdecorationsegmentamplitude}
        }
        {
        \pgfmathrand
        \pgfpointadd{\pgfqpoint{\pgfdecoratedinputsegmentremainingdistance}{0pt}}
                    {\pgfqpoint{-\pgfdecorationsegmentaspect
                     \pgfdecoratedinputsegmentremainingdistance}%
                               {\pgfmathresult\pgfdecorationsegmentamplitude}
                    }
        }
        {
        \pgfpointadd{\pgfpointdecoratedinputsegmentlast}{\pgfpoint{1pt}{1pt}}
        }
    }
    \state{final}{}
}

\definecolor{myred}{RGB}{215,25,28}
\definecolor{myorange}{RGB}{253,174,97}
\definecolor{myyellow}{RGB}{255,255,191}
\definecolor{mylightblue}{RGB}{171,217,233}
\definecolor{mydarkblue}{RGB}{44,123,182}

\newcommand{\rpm}{\raisebox{.2ex}{$\scriptstyle\pm$}}

\newtheorem{theorem}{Theorem}
\newtheorem{corollary}{Corollary}

\title{An Inverse Olympic Medal Tally Transformation for Optimal Lane-level Road Network Path Traversal}
\author{
  Dennis Luxen\\
  Frankfurt, Germany \\
  \texttt{ich@dennisluxen.de} \\
}

\begin{document}
\maketitle
\begin{abstract}
Lane-level traversal of (almost) arbitrary input paths is a common problem in the mapping industry.
This paper considers the problem of generating \emph{feasible} and maximally \emph{convenient} lane-level path traversals.
The presented approach exploits a graph transformation of an input path which is subsequentially explored by a multi-criteria search algorithm.
This approach is able to yield paths traversals that are guaranteed to obey lane crossing rules whenever possible, and minimize the number of legal, yet inconvenient lane crossings along any given input path.
\end{abstract}

\section{Introduction}
Services and devices that provide route planning and navigation fuctionality have become ubiquituous in daily life over the past two decades.
The purpose of route guidance is obviously to help a driver navigate a route, while lane-level guidance is a more recent improvement to navigation products.
The literature on how to generate lane-level guidance with certain quality guarantees is unfortuantely scarce.
Anecdotal knowledge has it that lane-level guidance in existing systems is computed greedily by independently inspecting individual intersections enhanced by some additional special-case handling heuristics.
Lane-level traversals through paths in a road network are an important ingredient into generating subsequent lane-level guidance information.
The method outlined in this paper is exact and gives quality guarantees.
Moreover, it also exhibits the same runtime asymptotics as widely used, yet inexact heuristics in industry.

\section{Related Work and Notation}
This paper follows the \emph{enhanced map (Emap)} model of {B\'{e}taille} and {Toledo-Moreo} \cite{bt-cemllvn-10} for representing lane-level connectivity in digital map data, which was later refined by Zhang et. al \cite{zagyc-llrnmgc-16}.
The authors investigate how to model and store map data sets that approximate vehicle trajectories across the lanes of road segments.
It is used here as follows:
The topology of the road network is modeled as a directed graph $G:=(V,E)$ with $\ell:E\rightarrow\mathbb{N}$ being the \emph{length (or cost) function} of the graph.
The graph is said to have $n:=\vert V\vert$ nodes (or vertices) and $m:=\vert E\vert$ edges.
Lane-level information augments the segment information in this model.
Nodes reflect intersections while edges resemble road segments as the connections in between.
Each edge $e\in E$ maps to a tuple $(L_e, Q_e), e\in E$ where $L_e$ is the set of lane information and $Q_e$ denotes the lane-independent, i.e. common, properties of the road segment.
E.g. this includes road class, center line geometry, street name, etc.
In turn, each $\lambda\in L_e$ is a tuple of lane-specific information such as lane shape, lateral lane sequence numbers, source/target intersections, among others.
As assumed by Chen et al. \cite{crf-hpllrmbvn-10} the maximum number of lanes along a segment is bounded by a small constant.
This paper makes three assumptions about this lane-level annotation.
First, given a turn of two connected and directed segments, the annotated information specifies which \emph{outgoing lanes} of a source edge connect to which \emph{incoming lanes} of the adjacent target edge.
Second, these relationships can be manifold, i.e. it may be possible to turn from one given source lane to more than one target lane.
And not all outgoing lanes must connect.
They could also just end at a segment.
Third, it is assumed that no map data set is perfect.
This is especially true for ever-changing data sets like OpenStreetMap \cite{rtc-uefmw-10,hw-ougsm-08}.
It's a popular and crowd-sourced data set with many concurrent streams of changes that may (or may not) introduce new (or fix any existing) inconsistencies.
In summary, a certain number of mapping data errors and inconsistencies are plainly unavoidable from a practical point of view in any such data set.
Any method using such data must arguably be able to gracefully handle errors.

A path $P:=\{e_0,e_1,\ldots,e_{n-1}\}$ is denoted by a sequence of edges of the road network graph.
Turn restrictions, i.e. invalid (or illegal) turns and turn sequences, are specified as \emph{forbidden} paths of two or more directed and linearly connected edges.
A lane-level traversal of a given path $P$ is a one to one association of road segment to one of the lanes it is augmented with.
It is called a \emph{feasible} traversal, if it does not include any forbidden lane-level turns and it's also called \emph{minimal} when it contains a minimal number of unwanted, yet legal lane-level turns only.
Unwanted turns are also referred to as \emph{inconvenient}.
Obviously, traversals can be pareto-optimal \cite{h-bpp-80}, i.e. not all traversals may be legal, or some may contain more unwanted maneuvers than others.

For the sake of this work, there are no assumptions on how input paths are generated.
They could be user-provided or computed by some method, or perhaps come from observations such as GPS trajectories.
The interested reader is referred to the excellent survey of Bast et al. \cite{bdgmpsww-rptn-16} for an overview on algorithmic results on how to generate optimal paths in a road network according to some travel mode specific metric.
Likewise, see Luxen and Vetter \cite{lv-rtrod-2011} for publicly and freely available implementations of some of these methods.
In principle, any generic shortest path methods should work with any network model having a sufficiently detailed modeling of possible turns, e.g. \cite{gv-errntc-11,w-mctrp-02}.

Anecdotal knowledge and industry experience shows that most if not all routing services don't route on a explicit road network model associating each lane with its own edge in the road network graph.
The reasons for this are likely twofold.
First, the blowup in computation and space is regarded as unwanted.
Second, it is also deemed as unnecessary:
The additional edges in the graph wouldn't model any connectivity in the road network that isn't already available and apparent from the model that associates a single edge with each road segment, and from sufficiently detailed turn restriction data.
It also explains why lane-level traversals are an ideal candidate for post-processing of any route-generating service -- given it is done in an exact way as the following  shows.

An \emph{Olympic Medal Tally} is a table of the medal counts of national teams which serves the purposes of identifying how well individual teams scored during a given years Olympic Games (amongst other potential reasons, e.g. \cite{hes-hinptomiun-10}).
Such a ranking sorts national teams lexicographically first by the number of gold medals.
Next, the number of silver medals for each team is considered.
And after that the number of bronze medals.
In other words, if the number of gold medals is the same, the number of silver medals serves as tie-breaker, which naturally extends to bronze.
Note that there are also other ways to rank teams.
The interested reader is referred to the relevant literature, e.g. \cite{ms-ciis-07}.

The remainder of this paper is structured as follows.
Section \ref{sec:lgg} analyses the worst case complexity of computing maximally feasible lane-level traversals of paths in a road network.
Section \ref{sec:llgc} introduces \emph{Lane-level Graph (LGGs)} transformations as a practical data structure to compute traversal of lane-level turns along a given path efficiently and with quality guarantees.
Section \ref{sec:applications} introduces an application of lane-level graphs to generate lane-based navigational guidance for drivers.
Section \ref{sec:cfw} summarizes the results and shows avenues of potential future work.

\section{Lane-Level Decision Complexity}\label{sec:lgg}

The layman's approach to generating lane-level path traversals is to inspect intersections independently, followed by a phase to greedily stitch them together with no or only a limited lookahead window.
It is arguably a crude, yet somewhat intuitive approach.
Unfortuantely, this fails to generate feasible of maximally convenient lane-level path traversals even for simple examples.
We introduce \emph{lane decision dependency}.
It describes that the decisions which lanes are legal to use may \emph{depend} on a possibly arbitrarily long chain of  preceding turns.
The example shown in Figure \ref{fig:example} visualizes this dependency on the right hand side:
\begin{figure}[b]
  \centering
  \begin{minipage}{0.45\textwidth}
      \begin{tikzpicture}[scale=1.25,decoration=penciline, decorate]

  \draw[decorate, line width=10,color=mydarkblue, opacity=0.75] (3.5,-1.0) -- (3.5, 0) -- (1.5,0) -- (1.5,1.0);

  \draw[<-,>=latex, thick] (0,0) -- (1.5,0);
  \draw[<-,>=latex, thick] (1.5,0) -- (3.5,0);
  \draw[<-,>=latex, thick] (3.5,0) -- (5.5,0);

  \draw[->,>=latex, thick] (1.5, 0) -- (1.5,1);
  \draw[<-,>=latex, thick] (3.5, 0) -- (3.5,-1);

  \node[anchor=south west] at (3.5,0) {$A$};
  \node[anchor=north east] at (1.5,0) {$B$};
  \node at (2.65,0.15) {$Q_e$};
  
  \node (A) at (1.5,0) [circle,fill=black,minimum size=2pt,scale=0.3] {};
  \node (B) at (3.5,0) [circle,fill=black,minimum size=2pt,scale=0.3] {};

  \node at (0,1) {};
  \node at (0,-1) {};
\end{tikzpicture}
  \end{minipage}\hfill
  \begin{minipage}{0.45\textwidth}
    \begin{tikzpicture}[scale=1.25,decoration=penciline, decorate]
  \draw[very thick] (1.25,1) -- (1.25,2);
  \draw[very thick] (1.75,1) -- (1.75,2);

  \draw[very thick] (3,0) -- (3,-1);
  \draw[very thick] (4,0) -- (4,-1);

  \draw[very thick] (0,1) -- (1.25,1);
  \draw[very thick] (1.75,1) -- (5,1);

  \draw[very thick] (0,0) -- (3,0);
  \draw[very thick] (4,0) -- (5,0);

  \draw[dashed,color=gray] (3.5,-0.25) -- (3.5,-1);

  \draw[solid,color=gray] (0.1, 0.5) -- (3.1, 0.5);
  \draw[solid,color=gray] (3.9, 0.5) -- (4.9, 0.5);

  \draw[decorate, line width=20,color=mydarkblue, opacity=0.75] (3.5, -1.0) -- (3.5, 0.5) -- (1.5,0.5) -- (1.5,2.0);

  \draw[rounded corners=5pt,->,>=latex, very thick, myred](3.25,-0.75) |- (2.75,0.25); 
  \draw[rounded corners=5pt,->,>=latex, very thick, myred](3.75,-0.75) |- (2.75,0.75); 
  \draw[rounded corners=5pt,->,>=latex, very thick, myred](2.5,0.75) -| (1.5,1.75); 
  \draw[rounded corners=5pt,->,>=latex, very thick, myred](2.5,0.25) -- (0.25,0.25); 

  \node[anchor=east] at (3.6,0.5) {$A$};
  \node[anchor=north east] at (1.6,0.9) {$B$};
  \node at (2.65,0.25) {$\lambda^1_e$};
  \node at (2.65,0.75) {$\lambda^2_e$};

\end{tikzpicture}
  \end{minipage}
  \caption{Left: Input road network (black) and route (blue); Right: detailed representation on the lane level.}
  \label{fig:example}
\end{figure}
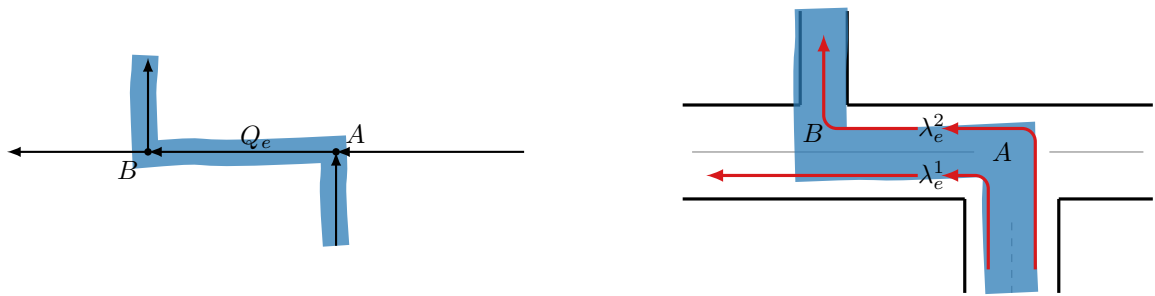
Not all lanes $\lambda^i_e\in Q_e, i\in \{1,2\}$ going into intersection $B$ connect to lanes following along the segment-level path.
The ability to follow along the path at intersection $B$ depends on the choice of lanes at the preceding intersection $A$.
This argument can be extended to length at most the number of segments in the entire path and it generalizes to convenient turns as well.
This is formalized in the following:
\begin{corollary}\label{cor:complexity}
  The complexity of lane-level decisions of a given path $P$ is linear in the number of segments in $P$.
\end{corollary}
In other words, one cannot derive any quality guarantees from an approach that conducts independent inspections of intersections and greedily accumulates the results.

\section{Lane-level Graph Transformation}\label{sec:llgc}
This Section gives the transformation of a path $P$ in $G$ into a \emph{lane-level graph} $G'_P$.
The resulting graph can be used to compute maximally feasible lane-level traversals that minimize unwanted maneuvers as much as possible.

At a high level, $G'_P$ resembles an augmented subset of $G$ in the vicinity of $P$ modeling lane-level connectivity plus some additional \emph{pods} to model whether a lane change is feasible (or not), and whether a turn from a lane of a source segment to a lane of a target segment is convenient (or not).
Note that the constructed graph has multi-criteria costs on its edges, i.e. costs are expressed as a tuple rather than a single numeric value.
The construction rules for a given path $P$ in $G$ are as follows:
\begin{enumerate}
  \setcounter{enumi}{-1}
  \item Create an empty graph $G`_P$ to be filled in subsequent steps.
  \item Add unconnected \textbf{new edges} $e_p$ in $G'_P$ for each of the lanes of segments $p\in P$ with cost $(0,0,0)$. Note that these edges do not share any vertices. 
  \item Connect all \textbf{feasible} turns from the out lanes of segment $\lambda^j_i$ to the in lanes $\lambda^j_{i+1}$ with a new edge $q$ and cost $(0,0,0)$. 
  \item Connect the edges $e$ corresponding to \textbf{unwanted, yet feasible maneuvers} with a new edge $q$ with cost $(0,1,0)$ at their beginning. 
  \item Connect all \textbf{infeasible, i.e. forbidden} turns from the out lanes of segment $p_i$ to the in lanes $p_{i+1}$ with cost vector $(1,0,0)$.
  \item Connect lateral lane change maneuvers from the beginning of $\lambda^i_p$ to the beginning of $\lambda^{i\rpm 1}_p$ with cost $(0,0,1)$.
\end{enumerate}
Figure \ref{fig:construction} gives a visual example of the transformation:
\begin{figure}[b]
  \centering
  \begin{tikzpicture}[scale=1.25,decoration=penciline, decorate]

  \node at (1,0) [rounded corners=3pt, minimum height=100pt, dashed,minimum width=40pt,draw, fill=mylightblue!20] {};
  \node at (3,0) [rounded corners=3pt, draw, fill=red!20, minimum height=100pt, dashed,minimum width=40pt,draw, fill=mylightblue!20] {};
  \node at (5,0) [rounded corners=3pt, draw, fill=red!20, minimum height=100pt, dashed,minimum width=40pt,draw, fill=mylightblue!20] {};

  \draw[thick,dashed] (-0.3,1) -- (0,1);
  \draw[->,>=latex, thick] (0,1) -- (0.7,1);
  \draw[thick,dashed] (-0.3,0) -- (0,0);
  \draw[->,>=latex, thick] (0,0) -- (0.7,0);
  \draw[thick,dashed] (-0.3,-1) -- (0,-1);
  \draw[->,>=latex, thick] (0,-1) -- (0.7,-1);

  \draw[->,>=latex, thick] (1.4,0.5) -- (2.7,0.5);
  \draw[->,>=latex, thick] (1.4,-0.5) -- (2.7,-0.5);

  \draw[->,>=latex, thick] (3.4,1) -- (4.7,1);
  \draw[->,>=latex, thick] (3.4,0.33) -- (4.7,0.33);
  \draw[->,>=latex, thick] (3.4,-0.33) -- (4.7,-0.33);
  \draw[->,>=latex, thick] (3.4,-1) -- (4.7,-1);

  \draw[thick] (5.3,1) -- (5.8,1);
  \draw[thick,dashed] (5.8,1) -- (6.2,1);
  \draw[thick] (5.3,0) -- (5.8,0);
  \draw[thick,dashed] (5.8,0) -- (6.2,0);

  \draw[color=gray!50,->,>=latex,very thick, dashed] (0.7,-1) to[bend left] (1.4,0.5);
  \draw[color=gray!50,->,>=latex,very thick, dashed] (0.7,0) to[bend right] (1.4,-0.5);
  \draw[color=gray!50,->,>=latex,very thick, dashed] (0.7,1) to[bend right] (1.4,-0.5);

  \draw[color=mydarkblue,->,>=latex,very thick] (0.7,1) to[bend right] (1.4,0.5);
  \draw[color=mydarkblue,->,>=latex,very thick] (0.7,0) to[bend left] (1.4,0.5);
  \draw[color=mydarkblue,->,>=latex,very thick] (0.7,-1) to[bend left] (1.4,-0.5);

  \draw[color=gray!50,->,>=latex,very thick, dashed] (2.7,0.5) to[bend right] (3.4,-1);
  \draw[color=gray!50,->,>=latex,very thick, dashed] (2.7,-0.5) to[bend right] (3.4,-1);
  \draw[color=gray!50,->,>=latex,very thick, dashed] (2.7,-0.5) to[bend left] (3.4,0.33);
  \draw[color=gray!50,->,>=latex,very thick, dashed] (2.7,-0.5) to[bend left] (3.4,1);

  \draw[color=mydarkblue,->,>=latex,very thick] (2.7,0.5) to[bend left] (3.4,1);
  \draw[color=mydarkblue,->,>=latex,very thick] (2.7,0.5) to[bend left] (3.4,0.33);
  \draw[color=mydarkblue,->,>=latex,very thick] (2.7,0.5) to[bend left] (3.4,-0.33);
  \draw[color=mydarkblue,->,>=latex,very thick] (2.7,-0.5) to[bend left] (3.4,-0.33);

  \draw[color=gray!50,->,>=latex,very thick, dashed] (4.7,1) to[bend right] (5.3,0);
  \draw[color=gray!50,->,>=latex,very thick, dashed] (4.7,-0.33) to[bend left] (5.3,1);
  \draw[color=gray!50,->,>=latex,very thick, dashed] (4.7,-1) to[bend left=10] (5.3,1);

  \draw[color=mydarkblue,->,>=latex,very thick] (4.7,1) to[bend left] (5.3,1);
  \draw[color=mydarkblue,->,>=latex,very thick] (4.7,0.33) to[bend right] (5.3,1);
  \draw[color=mydarkblue,->,>=latex,very thick] (4.7,0.33) to[bend right] (5.3,0);
  \draw[color=mydarkblue,->,>=latex,very thick] (4.7,-0.33) to[bend left] (5.3,0);
  \draw[color=mydarkblue,->,>=latex,very thick] (4.7,-1) to[bend right] (5.3,0);

  \draw[color=myorange,->,>=latex,very thick] (0.7,1) to[bend right=20] (0.7,0);
  \draw[color=myorange,->,>=latex,very thick] (0.7,0) to[bend right=20] (0.7,1);
  \draw[color=myorange,->,>=latex,very thick] (0.7,-1) to[bend right=20] (0.7,0);
  \draw[color=myorange,->,>=latex,very thick] (0.7,0) to[bend right=20] (0.7,-1);

  \draw[color=myorange,->,>=latex,very thick] (2.7,0.5) to[bend right=20] (2.7,-0.5);
  \draw[color=myorange,->,>=latex,very thick] (2.7,-0.5) to[bend right=20] (2.7,0.5);

  \draw[color=myorange,->,>=latex,very thick] (4.7,1) to[bend right=20] (4.7,0.33);
  \draw[color=myorange,->,>=latex,very thick] (4.7,0.33) to[bend right=20] (4.7,1);
  \draw[color=myorange,->,>=latex,very thick] (4.7,-0.33) to[bend right=20] (4.7,0.33);
  \draw[color=myorange,->,>=latex,very thick] (4.7,0.33) to[bend right=20] (4.7,-0.33);
  \draw[color=myorange,->,>=latex,very thick] (4.7,-1) to[bend right=20] (4.7,-0.33);
  \draw[color=myorange,->,>=latex,very thick] (4.7,-0.33) to[bend right=20] (4.7,-1);

  \node (vminus1) at (1,-2) {$v_{i-1}$};
  \node (v) at (3,-2) {$v_i$};
  \node (vplus1) at (5,-2) {$v_{i+1}$};

  \node (eminus2) at (0,-2) {$e_{i-2}$};
  \node (eminus1) at (2,-2) {$e_{i-1}$};
  \node (e) at (4,-2) {$e_i$};
  \node (eplus1) at (6,-2) {$e_{i+1}$};

  \draw[dashed,->,>=latex] (eminus2) -- (vminus1);
  \draw[dashed,->,>=latex] (vminus1) -- (eminus1);
  \draw[dashed,->,>=latex] (eminus1) -- (v);
  \draw[dashed,->,>=latex] (v) -- (e);
  \draw[dashed,->,>=latex] (e) -- (vplus1);
  \draw[dashed,->,>=latex] (vplus1) -- (eplus1);

  \node at (0.75,1.25) {$\mathcal{V}_{i-1}$};
  \node at (2.6,1.25) {$\mathcal{V}_{i}$};
  \node at (4.75,1.25) {$\mathcal{V}_{i+1}$};

  \node at (2,0.7) {\tiny{$\lambda^0_{p_{i-1}}$}};
  \node at (2,-0.3) {\tiny{$\lambda^1_{p_{i-1}}$}};

  \node at (4,1.2) {\tiny{$\lambda^0_{p_{i}}$}};
  \node at (4,0.53) {\tiny{$\lambda^1_{p_{i}}$}};
  \node at (4,0.075) {\tiny{$\vdots$}};
  \node at (4,-0.79) {\tiny{$\lambda^n_{p_{i}}$}};
\end{tikzpicture}
  \caption{Constructed lane-level graph with pods (dashed rectangles), feasible turns (black), feasible yet unwanted connections (blue), as well as lane change maneuvers (orange), infeasible maneuvers (dashed lines) for an example path input path $P:=(\ldots, e_{i-1}, e_i, e_{i+1}, \ldots)$ in $G$.}
  \label{fig:construction}
\end{figure}
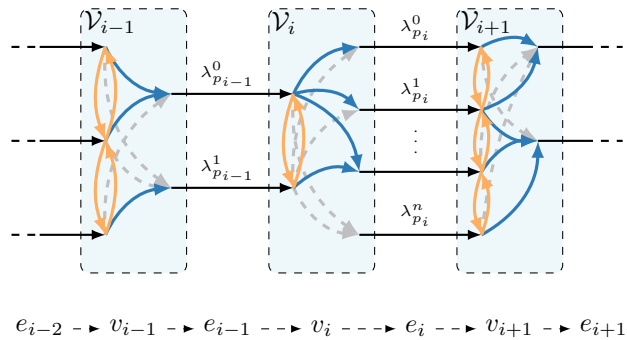
The lane segments connect to \emph{pods} (dashed rectangles) that correspond to the turns from one segment to the next.
Allowed lane-level turns are reprented by solid arcs connecting lane edges within a pod from left to right.
Lateral lane change maneuvers connect lanes at the end of a segment (orange).
Maneuvers that are feasible, yet unwanted are given in blue.
Infeasible maneuvers (dashed) are added to make sure that there always exists a path from sources to sinks to guarantee connectedness.
\footnote{Note that the described graph transformation is a bijective function, because collapsing pods into a single vertex and removing any resulting duplicate edges yields the original input path $P$ again.}

The number of pods is linear in the number of segments in $P$, while the number of lanes define the internal size of a pod.
From a practical point of view the maximum number of lanes is bounded by a small constant as most road segments only have a single lane per direction.
The widest road in the world is believed to be the the Katy Freeway, or Interstate 10 close to Houston, Texas, USA with 26 lanes\footnote{cf. \url{https://en.wikipedia.org/wiki/Interstate_10_in_Texas}}.
The internal size of a pod is bounded by
\begin{equation}\label{eq:podsize}
  k^2+2\cdot (k-1) = \mathcal{O}\left(n^2\right)
\end{equation}
connections it contains.
The quadratic term in Equation \ref{eq:podsize} reflects the maximum number of edges in bipartite graphs, while the other term corresponds to maximum number of lane change maneuvers.
As mentioned above it is assumed that in reality $k$ is bounded by a small constant.
Hence, the size of a pod is assumed to be bounded by a constant and as a result the size of a lane-level graph is also assumed to be linear in the size of the input path for real-world data.

A path in $G'_P$ from the source pod to the target pod corresponds to a lane-level traversal.
Note again that costs in the graph are expressed as a tuple.
A best path that is feasible, and contains neither any unwanted maneuvers nor requires any lateral lane changes obviously has cost $(0,0,0)$, when costs are accumulated component-wise.
The first component gives the number of infeasible turns, and the second one gives the number of unwanted, yet feasible turns. The third component gives the number of lateral lane changes.
Also, note that the cost tuples of paths relate to the olympic medal tally with its counts of medals of different kinds.
A path with the most number of \grqq{} gold\grqq{} medals contains the most number of infeasible lane-based turns.
Likewise, the path with the most number of \grqq{}silver\grqq{} medals contains the most number of unwanted -- yet legal -- turns, etc.
In that sense, the goal of finding a maximum feasible lane-level path traversal with a minimal number of unwanted lane maneuvers corresponds to the top contender on an \emph{inverse} Olympic Medal Tally of all possible lane-level paths corresponding to input path $P$.

It is easy to see that a lane-level graph allows to generate an exact lane-level traversal with quality guarantees in polynomial time by computing a multi-criteria least-cost path from the sources of $G'_P$ to the sinks.
It suffices to zero-initialize cost tuples of source lanes in the priority queue that guides the search.
This gives the following corollary:

\begin{corollary}[Correctness and Exactness]
Running a queue-initialized multi-source, multi-criteria least-cost path computation with a lexicographic comparison function generates a maximally feasible lane-level path traversal of $P$ having a minimal number of unwanted maneuvers and lateral lane changes.
\end{corollary}

The worst-case (time and space) complexity of generating correct lane-level guidance is obviously determined by the running time of a shortest path query on $G':=(V', E')$.
This is known to be $O(\vert E'\vert \log \vert E'\vert )$ using Dijkstra's seminal algorithm \cite{d-ntpcwg-59} as $V\in O(\vert E'\vert\cdot k)$.
The logarithmic factor is due to the fact that efficient implementations of Dijkstra's method use a min-priority queue to generate the order in which the explored vertices of $G'_P$ are settled.
While this may already be efficient enough from a practical point of view, the theoretical runtime bound can be improved upon by getting rid of this logarithmic factor.
First, we give the runtime theorem, and then sketch its proof:
\begin{theorem}[Runtime]
  Optimal lane-level traversals of paths $P$ can be computed in time $O(\vert G'_P\vert)$, i.e. in time linear in the size of its corresponding lane-level graph.
\end{theorem}
Linear runtime becomes possible by removing the need to determine the order in which nodes in $G'_P$ are settled by the search -- which directly follows from the construction of $G'_P$.
Note that the following is an inductive argument.
\begin{proof}
The weights at vertices of the first pod $v\in\mathcal{V}_{0}$ are obviously zero-initialized and thus minimal by definition.
Once weights for $\mathcal{V}_i$ are set, weights for $\mathcal{V}_{i+1}$ can be determined.
All vertices $v\in\mathcal{V}_{i+1}$ can be settled after vertices $\mathcal{V}_i$ are settled by scanning the directed edges in $G'_P$ from $\mathcal{V}_i\rightarrow\mathcal{V}_{i+1}$ (plus potentially pod-internal links of $\mathcal{V}_{i+1}$) and noting minima weights at all $v\in\mathcal{V}_{i+1}$.
\end{proof}
The lane-level graph can be traversed pod by pod and could even be generated on the fly.
This can also be seen in Figure \ref{fig:construction}:
Directed edges exist only between the (dashed rectangle) pods in the same directions.
The order in which to settle internal nodes of a pod can be computed in sorting time $O(k\cdot\log k)$.
Again, $k$ is a small constant and for practical purposes, the order could be computed by a fixed-size sorting network.
In other words, the weights for nodes in a pod corresponding to a vertex $v_{i}\in P$ can be found sequentially by scanning and relaxing outgoing edges from the predecessor vertex $v_{i-1}$.
Note the similarity of this kind of graph exploration to that of a linear sweep across the input path $P$.

\section{Application: Lane-level Guidance}\label{sec:applications}
This Section highlights a natural application of the above method in practice along with some pointers into the related literature.
The idea of providing guidance to motorists along their routes by electronic means dates back several decades already as outlined by Rosen et al \cite{rmf-ergshv-70}:
The aim of providing guidance is to give \glqq the maneuvers necessary the driver should make at the upcoming intersection or interchange to reach his destination optimally\grqq.
Lovelace et al. \cite{lhm-egrdfue-99} describe that higher guidance quality is correlated with mentioning more details such as more deliberate segment as well as turn descriptions.
Lane-level guidance must navigate a driver through complex situations in the road network that depend on being in the right lane at the right time.
Generally speaking, the advantages are obvious and can have a positive social impact:
Guidance not only eases the uncertainty of drivers at decision points along a route, it also avoids excessive travel in the road network.

The \emph{guidance} $G_P$ of a path $P$ describes additional attributes that are attached to a subset $V_{P_G}\subseteq P_G$ of a paths vertices.
At times these vertices are called decision points and represent the locations along the route where guidance shall be issued to a driver.
The literature on the algorithmic generation of guidance for a path in a road network is scarce, unfortunately.
For segment-based guidance Luxen \cite{l-frgttig-13} introduced the concept of \emph{turn instruction graphs (TIGs)} to identify decision points in a generic way that allows applying so-called one-size-fits-all rules as well as user-specific ones.
Guidance is generated by performing a graph transformation and by subsequently contracting edges of the resulting directed, acyclic graph repeatedly until the generated overlay is \emph{minimal} according to the aforementioned set of rules.

Lane-level guidance is said to be \emph{feasible} if none of the generated instructions on a source lane lead a target lane from which the next turn is impossible.
The subject of lane-level guidance has only seen a few related results -- mostly focusing on presentation of already-generated guidance.
For example, Lee at al. \cite{lykpk-dllgsivars-15} investigate an augmented reality system with lane-level information display.
Cao et al. \cite{clzz-aarbvn-18} describe a navigation system that relies on lane-precise localization and guidance generation.

The main observation here is that the foundational task of lane-level guidance is to find a valid lane-level traversal of a given input path.
Constructing a lane-level graph and running a least-cost path query as explained in prior sections from the sources of $\mathcal{V}_0$ to the sinks of $\mathcal{V}_{n-1}$ yields a sequence of lateral lane numbers.
These are the lanes that have to be taken by a driver along the route.
Routes that minimize the number of lane changes and also have best-possible feasibility are obviously superior to others.

\section{Conclusion and Future Work}\label{sec:cfw}
This paper presents a graph transformation followed by a multi-criteria least-cost path query to evaluate the lane-level traversal of given routes.
It yields traversals on the lane-level that exhibit quality guarantees.
Traversals are maximally feasible and likewise avoid inconvenient situations and even lateral lane changes as much as possible.

The outlined methods lends itself to an extensive future experimental evaluation including comparative analysis of distinct map data sets with dozens of hundreds of millions of road segments and an even greater number of routes/paths.
This should also encompass a comparison of the opportunistic solution with a fixed window size on real-world data sets.

An interesting avenue for future research could be to use a cost function with non-integer component values and to evaluate lane-level traversals in a fuzzy evaluation scheme similar to the one outlined by Delling et al. \cite{ddpww-cmjp-13} to prune sets of pareto optimal transit journeys before presenting them to a user.
The latter of which could be an interesting input into the ranking of alternate routes, e.g. \cite{ls-csarrn-15}, of routes in road networks.

\vfill
\pagebreak
\bibliographystyle{plain}
\bibliography{mybibliography}
\end{document}